\documentclass[prl,reprint,twocolumn,amsmath,preprintnumbers,amssymb,superscriptaddress]{revtex4-1}

\usepackage{times}
\usepackage{array}
\usepackage{graphicx}
\usepackage{color}
\usepackage{hyperref}
\usepackage{braket}
\usepackage{multirow}
\usepackage{indentfirst}
\usepackage{srcltx}
\usepackage[FIGTOPCAP,raggedright,nooneline,bf,footnotesize]{subfigure}
\usepackage{paralist}
\usepackage[usenames,dvipsnames]{pstricks}
\usepackage{epsfig}
\usepackage{overpic}
\newcommand{\figref}[1]{Fig.~\ref{#1}}

\begin{document}
%\title{Time-resolved double-slit interference pattern measurement with entangled photons }
\title{Time-resolved double-slit experiment with entangled photons}
\author{Piotr Kolenderski}
\email{kolenderski@fizyka.umk.pl}
\affiliation{Institute for Quantum Computing and Department of Physics and Astronomy, University of Waterloo, Waterloo, Ontario, N2L 3G1, Canada}
\affiliation{Institute of Physics, Faculty of Physics, Astronomy and Informatics, Nicolaus Copernicus University, Grudziadzka 5, 87-100 Torun, Poland}

\author{Carmelo Scarcella}
\affiliation{Politecnico di Milano,  Dipartimento di Elettronica, Informazione e Bioingegneria, Piazza Leonardo da Vinci 32, I-20133 Milano, Italy}

\author{Kelsey D. Johnsen}
\author{Deny R. Hamel}
\author{Catherine Holloway}
\affiliation{Institute for Quantum Computing and Department of Physics and Astronomy, University of Waterloo, Waterloo, Ontario, N2L 3G1, Canada}

\author{Lynden K. Shalm}
\affiliation{Institute for Quantum Computing and Department of Physics and Astronomy, University of Waterloo, Waterloo, Ontario, N2L 3G1, Canada}
\affiliation{National Institute of Standards and Technology (NIST), 325 Broadway, Boulder, CO 80305, USA}

\author{Simone Tisa}
\affiliation{Micro Photon Device S.r.l., Via Stradivari 4, I-39100 Bolzano, Italy}

\author{Alberto Tosi}
\affiliation{Politecnico di Milano,  Dipartimento di Elettronica, Informazione e Bioingegneria, Piazza Leonardo da Vinci 32, I-20133 Milano, Italy}
\author{Kevin J. Resch}
\author{Thomas Jennewein}
\affiliation{Institute for Quantum Computing and Department of Physics and Astronomy, University of Waterloo, Waterloo, Ontario, N2L 3G1, Canada}

\maketitle
% Frabboni2010,
\textbf{The double-slit experiment strikingly demonstrates the wave-particle duality of  quantum objects. In this famous experiment, particles pass one-by-one through a pair of slits and are detected on a distant screen. A distinct wave-like pattern emerges after many discrete particle impacts as if each particle is passing through both slits and interfering with itself. While the direct event-by-event buildup of this interference pattern has been observed for massive particles such as electrons \cite{Tonomura1989, Frabboni2011,Bach2013}, neutrons \cite{Zeilinger1988}, atoms \cite{Carnal1991} and molecules \cite{Arndt1999, Brezger2002}, it has not yet been measured for massless particles like photons. Here we present a temporally- and spatially-resolved measurement of the double-slit interference pattern using single photons.  We send single photons through a birefringent double-slit apparatus and use a linear array of single-photon detectors to observe the developing interference pattern. The analysis of the buildup allows us to compare quantum mechanics and the corpuscular model described in Ref.~ \cite{Jin2010a}, which aims to explain the mystery of single-particle interference. Finally, we send one photon from an entangled pair through our double-slit setup and show the dependence of the resulting interference pattern on the twin photon's measured state. Our results provide new insight into the dynamics of the buildup process in the double-slit experiment, and can be used as a valuable resource in quantum information applications.
}

Interference pattern measurements using individual photons have been carried out with relatively slow exposing charge-coupled device (CCD) cameras \cite{Garcia2002,Jacques2005,Fickler2012} and by scanning a single-photon detector through a detection plane \cite{Zeilinger2005a}. However, CCD cameras do not resolve the impact of individual photons, and scanning single-photon detectors cannot simultaneously record full spatial and temporal information. In our setup, we use an array of 32 single-photon avalanche detectors (SPAD) \cite{Guerrieri2010a,Tisa2007} as a detection ``screen'' for our double-slit setup. Using this SPAD array in our interference setup, we are able to observe the buildup of the double-slit interference pattern with high resolution in both space and time.

\begin{figure}[t]
\centering
{\begin{overpic}[width=\columnwidth,trim = 10mm 8mm 0mm 20mm, clip]{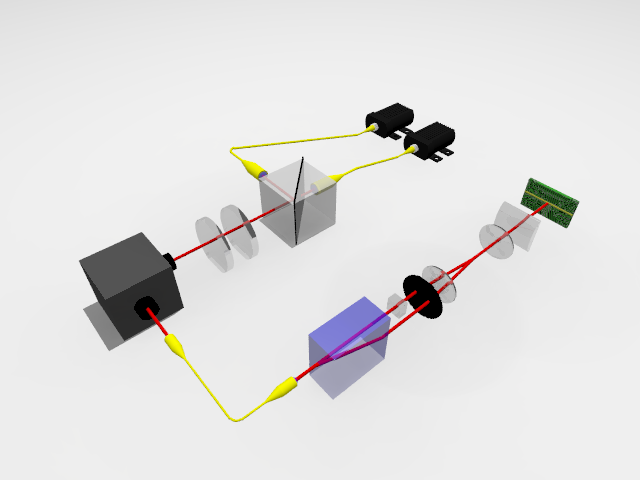}
%\put(1,14){\scriptsize \textbf{(b)} SPAD array}
\put(69,1.5){{\includegraphics[width=0.3\columnwidth]{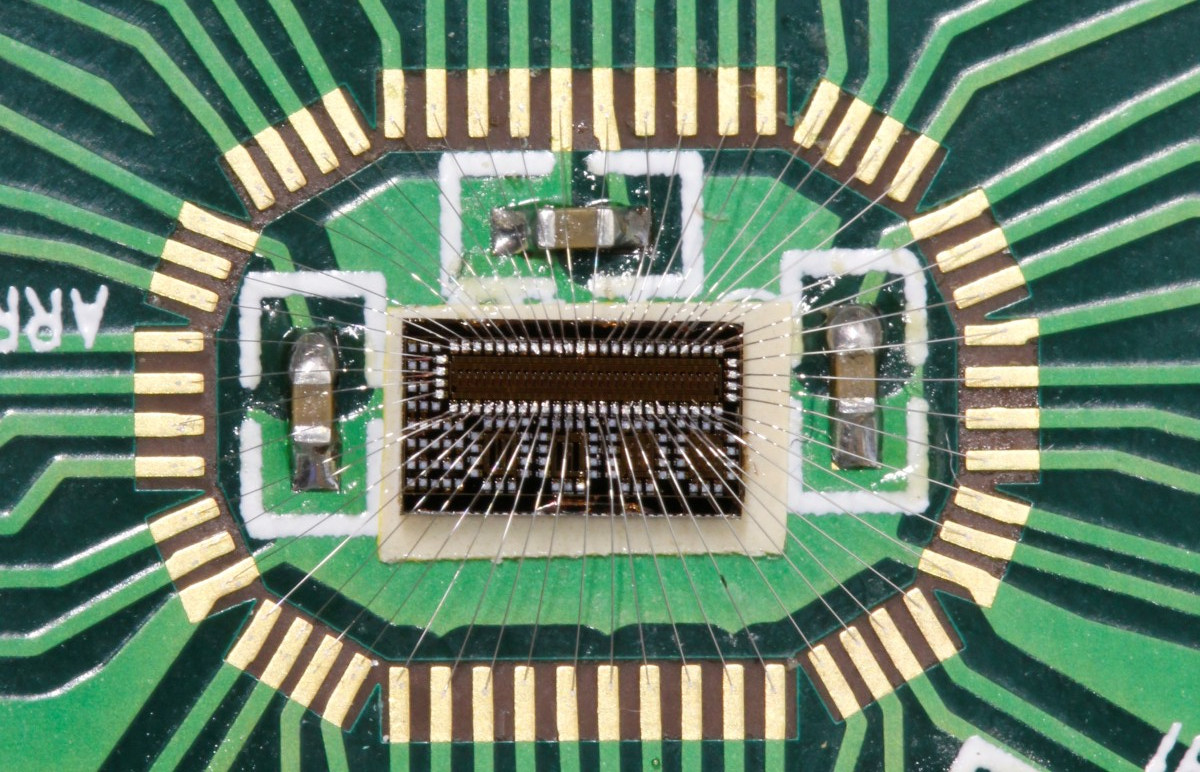}}}
\put(88.05,46.75){\scriptsize SPAD}
\put(88.08,44){\scriptsize Array}
\put(64,55){\scriptsize D1}
\put(71,51.5){\scriptsize D2}
\put(48,32.25){\scriptsize PBS}
\put(31,25){\scriptsize HWP}
\put(36.5,29){\scriptsize QWP}
\put(1,23){\scriptsize Sagnac}
\put(1.4,20){\scriptsize source}
\put(42,23){\scriptsize Calcite}
\put(53,26){\scriptsize CC}
\put(23,41){\scriptsize 776 nm}
\put(16,8){ \scriptsize 842 nm}
\put(58,28){\scriptsize P}
\put(71,38.5){\scriptsize L3}
\put(68,36){\scriptsize L2}
\put(60,30.25){\scriptsize L1}
\end{overpic}}
\caption{Experimental setup. The Sagnac-type source produces polarization entangled photon pairs. One photon is coupled into single-mode fiber. A birefringent calcite crystal displaces photons with horizontal polarization, and a crystal (CC) compensates for path length difference. A polarizer (P) erases any distinguishing information about the photons. Two lenses (L1 and L2) determine beam size, and a third lens (L3) focuses the beam vertically onto the SPAD detectors. The other photon is sent through a polarization analyser consisting of a half wave plate (HWP), quarter-wave plate (QWP) and polarizing beamsplitter (PBS). It is then coupled into one of two single-mode fibres connected to detectors (D1 and D2). The inset shows a photo of the 32-pixel SPAD array \cite{Guerrieri2010a,Tisa2007}.   }
\label{fig:setup:spdc}
\end{figure}
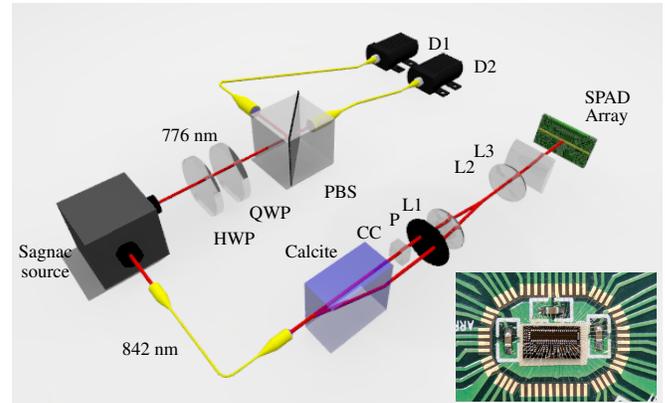

%\cite{Kim2006,Fedrizzi2007}
Our experimental setup, shown in \figref{fig:setup:spdc}, uses a Sagnac-type source \cite{Hamel2010} producing photon pairs, a vertically-polarised photon at $842$ nm and a horizontally-polarised photon at $776$ nm, via the nonlinear process of spontaneous parametric downconversion (SPDC) \cite{Hubel2010}.  The $776$ nm photon passes through a series of waveplates and a polarizing beamsplitter. It is detected by D2 and used to herald the presence of the $842$ nm photon \cite{Rarity1987}.
The $842$ nm photon is coupled into a single-mode fibre, and a polarization controller (not shown) prepares the state in an equal superposition of horizontal (H) and vertical (V) polarizations. This is then outcoupled, resulting in a free-space Gaussian spatial mode with a waist of 1.3 mm. This beam is collimated and sent to a polarization-based double slit composed of a calcite beam displacer. The birefringence of this crystal results in the displacement of horizontally polarized photons by $3.68$ mm with respect to the vertically polarized photons. The beam displacer maps the polarization state of a photon into a spatial state, which is encoded in its path. These two paths are analogous to a double-slit apparatus. They are orthogonally polarized and thus carry distinguishing information, which is erased by a polarizer set at $45$ degrees. A compensating crystal (CC) is placed after the beam displacer to make the two path lengths equal, and a series of lenses maps the interference pattern onto the SPAD array.

\begin{figure}[t]
\centering
\begin{tabular}{c c}
\multirow{3}{*}{\subfigure[$N=200$, $t=115$ ms]{\includegraphics[width=0.47\columnwidth]{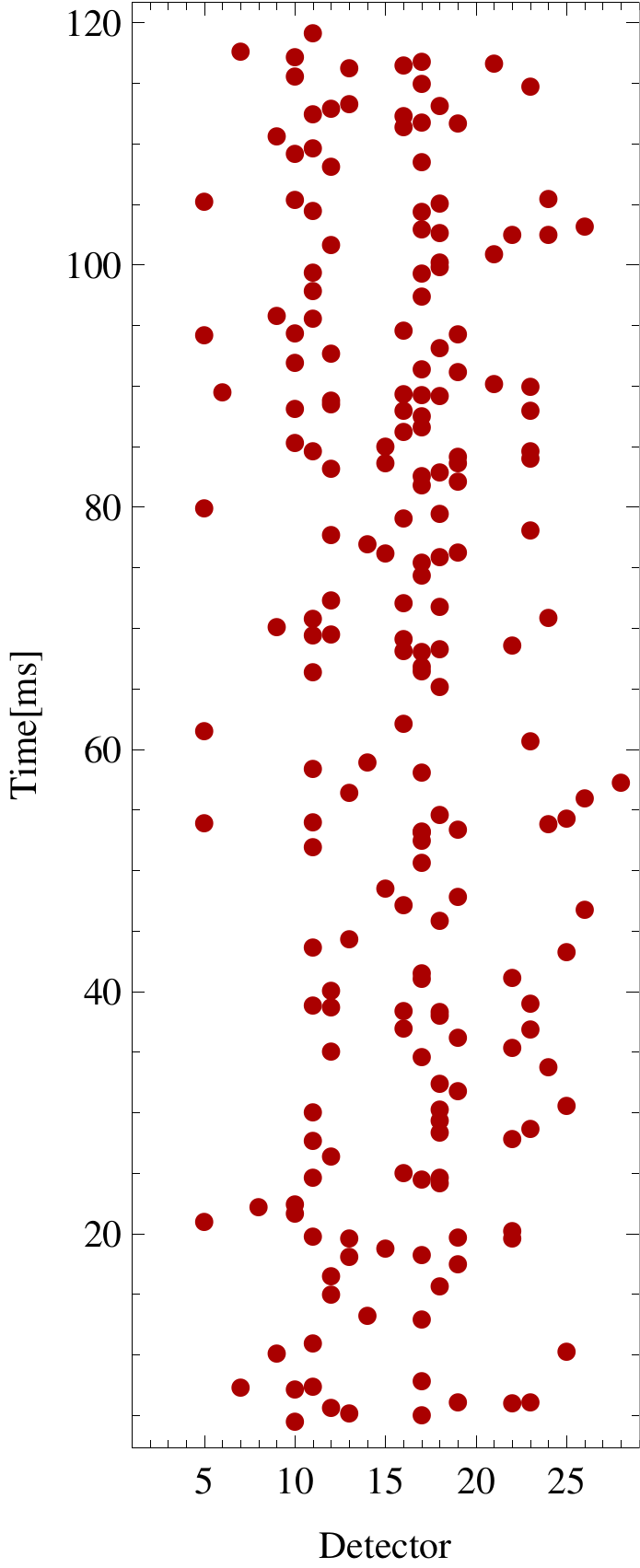}}}  & \vspace{2.8cm}
\multirow{2}{*}{\subfigure[$N=2000$, $t=1.2$ s]{\includegraphics[width=0.48\columnwidth]{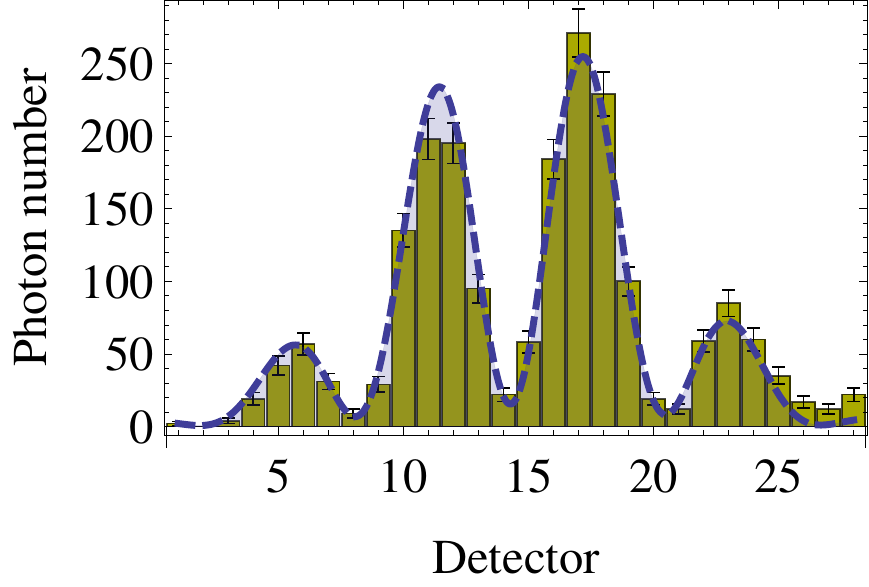}}}  \\&
\subfigure[$N=200$, $t=115$ ms]{\hspace{.2cm}\includegraphics[width=0.46\columnwidth]{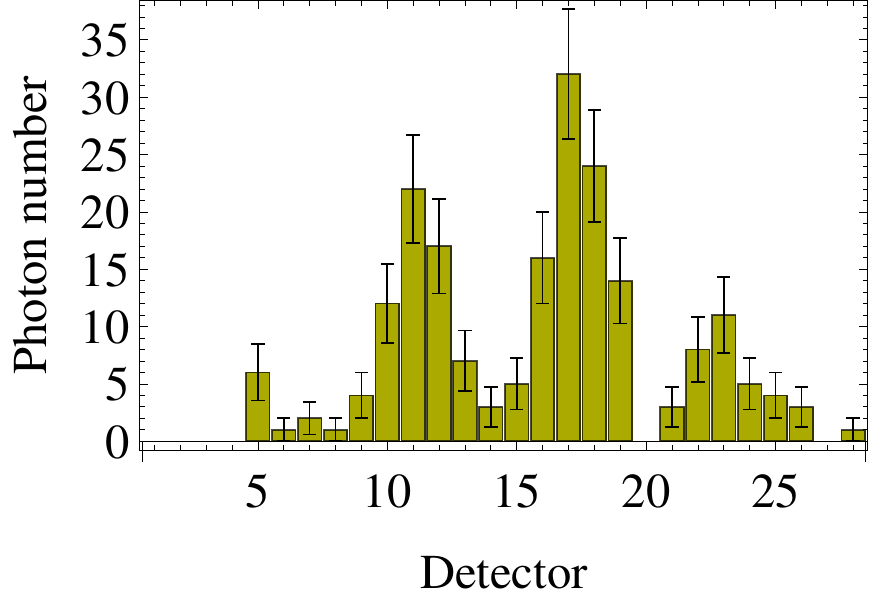}}\\&
\subfigure[$N=20$, $t=13$ ms]{\hspace{.2cm}\includegraphics[width=0.46\columnwidth]{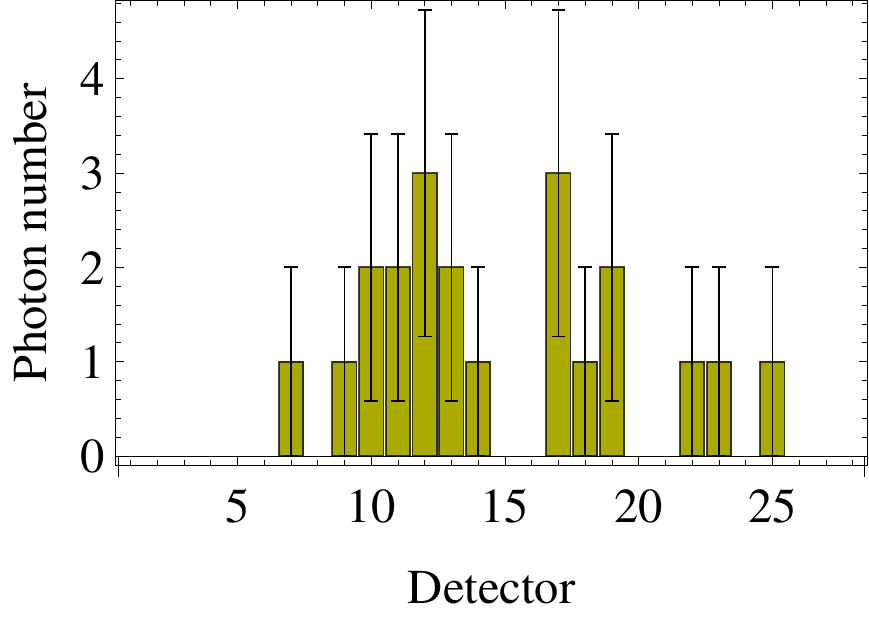}} \\
&\\
%\multicolumn{2}{c}{\subfigure[\ $g^{(2)}$]{\includegraphics[width=0.8\columnwidth]{fig-g2-hist-array}}}
 
%\subfigure[]{\includegraphics[width=0.45\columnwidth]{fig-g2-hist-D1-array}}
%\subfigure[\ Interference quality]{\includegraphics[width=0.45\columnwidth]{bd-interf-rotA-ch15-R2-belt}}
\end{tabular}
\caption{Interference pattern buildup. Panel (A) shows first $200$ heralded counts in time, and panels (B-D) depict the statistics of the first $20$, $200$ and $2000$ heralded detections.}
\label{fig:reel:spdc}
\end{figure}

Each of the 32 detectors in the SPAD array records the arrival time of single photons with a  timing uncertainty of about $150$ ps, which is the combined timing jitter of the detectors and time tagging logic. \figref{fig:reel:spdc}(A) shows the arrival times of the first $200$ detection events passing through the slits. The accumulation of these events results in an interference pattern, as shown in \figref{fig:reel:spdc}(B-D). After the detection of $2000$ photons, the interference pattern becomes very clear, with a visibility of $93 \pm 2$\%. This visibility is not perfect as a result of inexact compensation of the two path lengths. A movie and additional measurements using a coherent source can be found in Supplementary Information.

%\com{this next paragraph with the autocorrelation confuses me. It only looks at 200 events and then tries to conclude some kind of statistical significance. This just tells us that the events are uncorrelated (which would be true for a coherent source) The really interest is in second order correlations like a g2 which we do not have enough data for. This section needs to be explained better} 
%In addition to looking at the relative time differences between successive detections at the SPAD array, we analyse the temporal and spatial characteristics of consecutive photon detections (see Supplementary information for details) in order to determine whether or not they are correlated. The average of both the spatial and temporal auto-correlation functions was calculated to be $0 \pm 0.1$. This clearly shows randomness in both time and space.

\begin{figure}[ht]
\begin{tabular}{c c}
%\subfigure[]{\includegraphics[width=0.49\columnwidth]{fig-likelihood-test-calcite}} &
%\subfigure[]{\includegraphics[width=0.43\columnwidth]{bd-interf-rotA-ch15-R2-QMvsCorp}}\\
%\subfigure[]{\includegraphics[width=0.49\columnwidth]{fig-likelihood-test-calcite-1k}} &
%\subfigure[]{\includegraphics[width=0.43\columnwidth]{bd-interf-rotA-ch15-R2-QMvsCorp-1k}}\\
\subfigure[\ coefficeint of determination ]{\includegraphics[width=0.44\columnwidth]{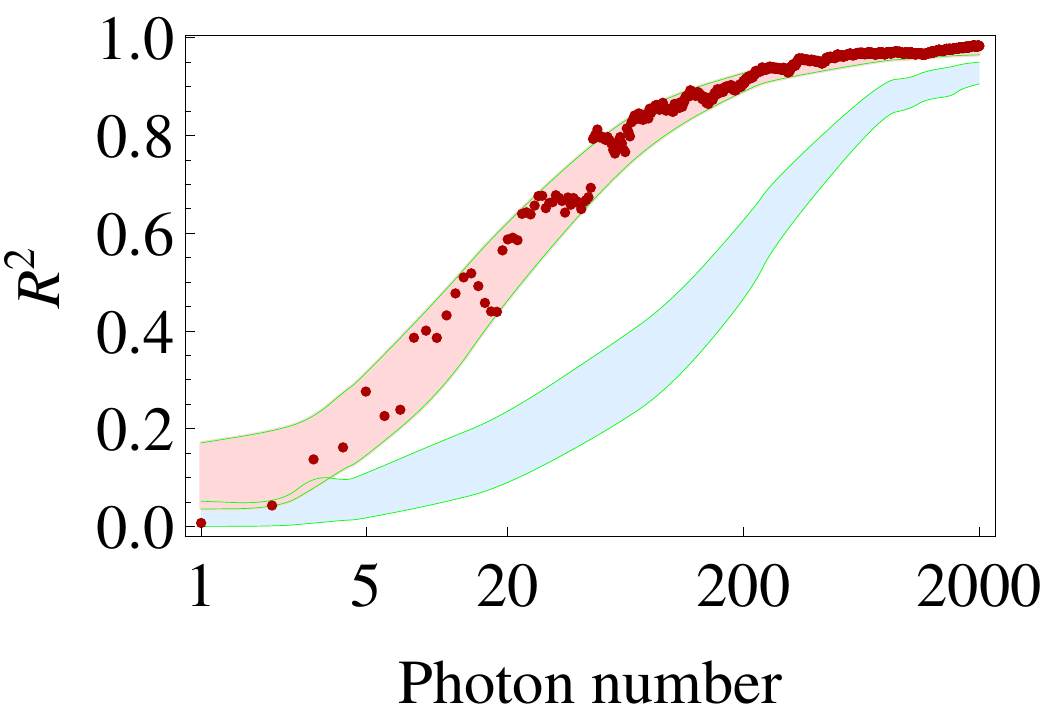}} &
\subfigure[\ likelihood ratio test]{\includegraphics[width=0.46\columnwidth]{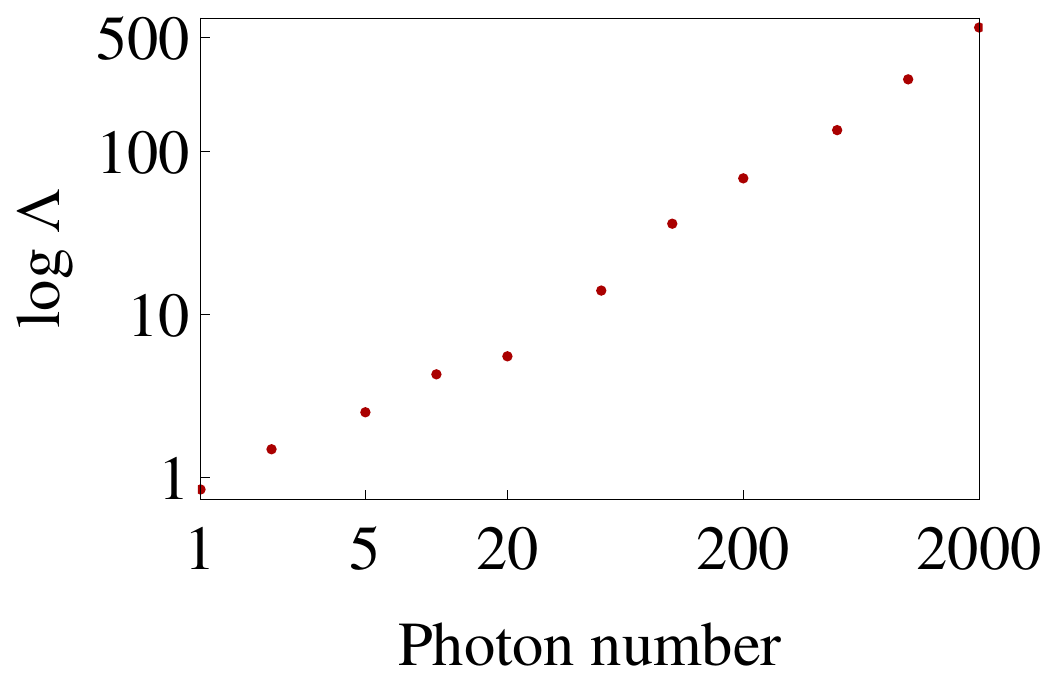}}\\
\end{tabular}
\caption{Statistical tests. (A) Coefficient of determination. For a given photon number, the statistics of $R^2$ is generated after $10^5$ numerical Monte Carlo simulations for the corpuscular and quantum mechanical models. The red (blue) belt shows $50$\% of the most frequent values of $R^2$ for the case of the quantum mechanical (corpuscular) model. The red points are the values of $R^2$ for experimental data. (B) Likelihood ratio test. The smallest likelihood ratio value is $\log \Lambda = 0.83$, which shows that quantum mechanics is a better indicator of the behaviour seen in nature.}
\label{fig:cod:spdc}
\end{figure}

%event-by-event
Our ability to accurately measure the arrival times of photons allows us to test the predictions of an alternative corpuscular theory, designed to explain the phenomenon of interference without wave-particle duality \cite{Jin2010a}. In this theory, detectors are modelled as deterministic learning machines, which are able to reproduce the interference pattern after many photon detections. The detectors' internal states update after each photon detection (see discussion in Supplementary Information), improving their knowledge of the pattern. 

Using two statistical methods and the measured buildup of the interference pattern, we examine the predictions of this corpuscular theory and quantum mechanics. The coefficient of determination \cite{Zwillinger1995}, $R^{2}$, allows us to evaluate how well each model predicts the final interference pattern with increasing detection number, while the likelihood ratio test \cite{Casella2001}, $\Lambda$, allows us to compare the two models. 

% R2
%First, we fit the photon statistics acquired after detection of $98000$ photons to the quantum mechanical model using intensity, transverse shift and magnification. The remaining setup parameters are fixed to measured values. The resulting coefficient $R^2$ is $0.99$, which tells us that the quantum mechanical model predicts the final pattern very well. First, we simulate the statistics of detected photons using the corpuscular model  \cite{Jin2010a} and compare this to the final interference pattern predticted by wave mechanics. This results in a maximum goodness-of-fit coefficient $0.98$. This indicates that the quantum mechanical and the corpuscular model statistics very closely predict the wave mechanical intereference pattern in the limit of high photon numbers.

We begin by calculating the coefficient of determination, $R^2$, to see how quickly the measured data, the corpuscular model and quantum mechanics each reproduce the final interference pattern. This pattern is derived from classical wave mechanics, and intensity is used as the only fit parameter (see analysis of interference pattern in Supplementary Information). Our experimental data gives us $R^2=0.96$ after $190 \pm 5$ detections, as shown in \figref{fig:cod:spdc}(A). This tells us that the interference pattern is clearly visible after only $190$ detection, which we then use as a reference for comparison with the two models. Next, we use the Monte Carlo method to run $10^5$ numerical simulations of $1\dots 2000$ photon detections for quantum mechanics and the corpuscular model. The statistics of $R^2$ for these simulations are shown in \figref{fig:cod:spdc}(A). Although both quantum mechanics and the corpuscular model eventually predict the final pattern very well, they require $200 \pm 5$ and $1000 \pm 10$ photons, respectively, to achieve $R^2=0.96$. While it is clear that these statistics for the quantum mechanical simulations and experimental data have similar trends, the coefficient of determination cannot conclusively say which model is better.

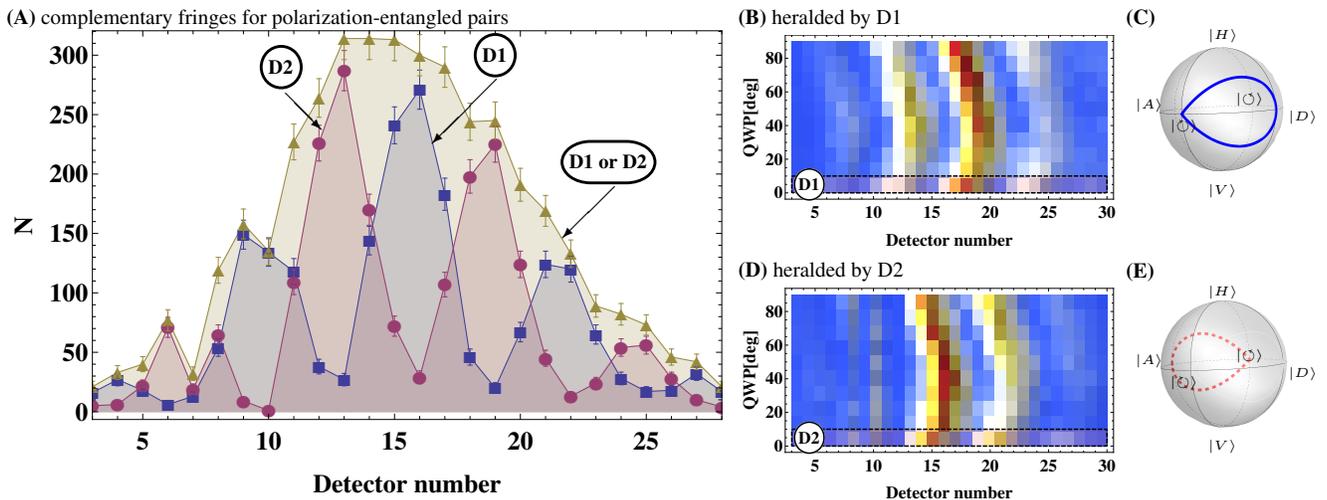
\begin{figure*}[ht]
\psset{cornersize=absolute,linearc=.5\baselineskip}
\centering
\begin{tabular}{c c c}
\multirow{2}{*}{\subfigure[\ complementary fringes for polarization-entangled pairs ]{\begin{overpic}[width=1.1\columnwidth]{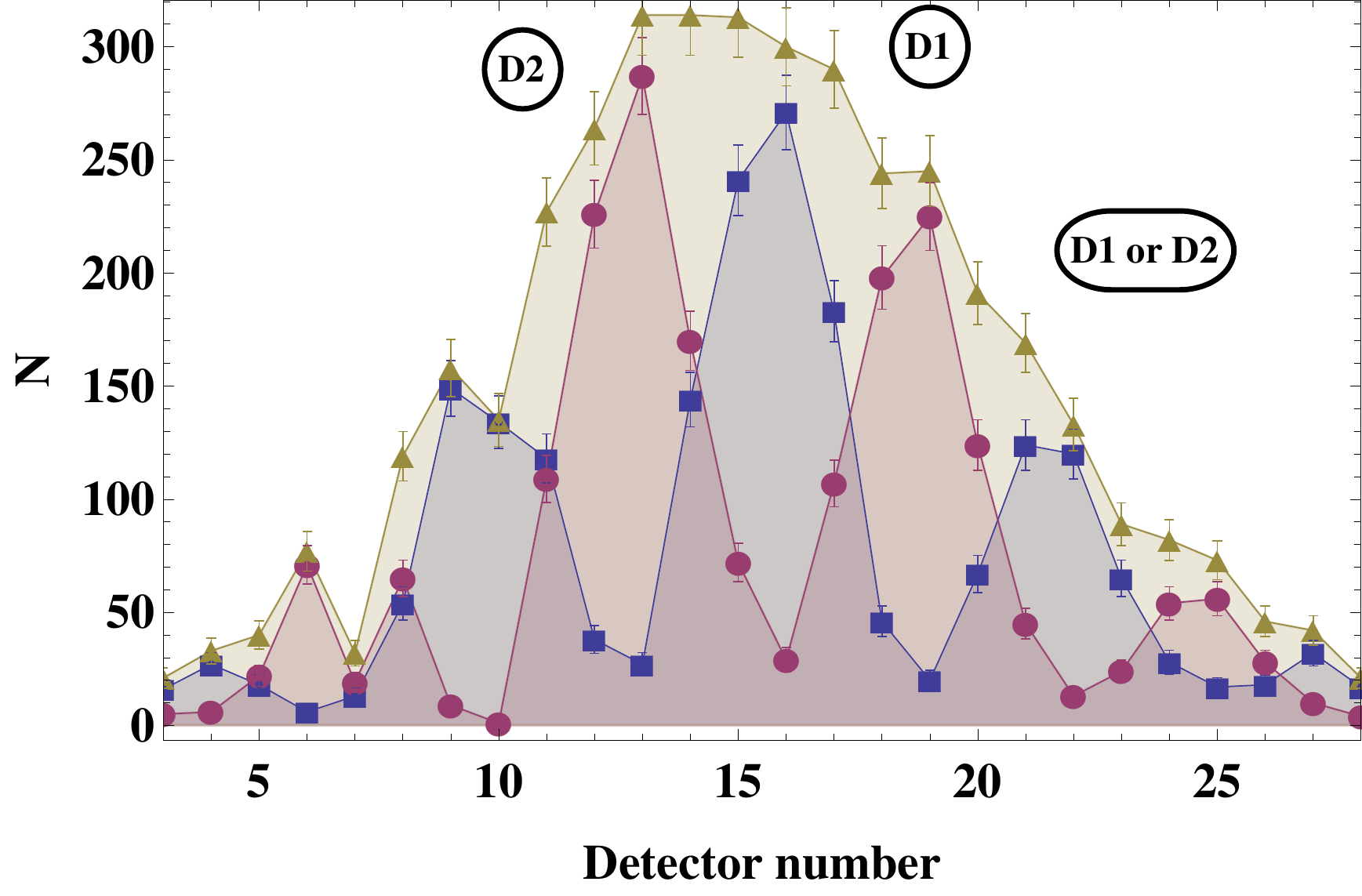}
\put(68.5,59){\vector(-1,-1){9}}
\put(39,57){\vector(1,-1){5}}
\put(85,44){\vector(-1,-1){7.25}}
\end{overpic}}} \vspace{3cm}&
\multirow{2}{*}{\subfigure[\ heralded by D1]{\begin{overpic}[width=0.58\columnwidth]{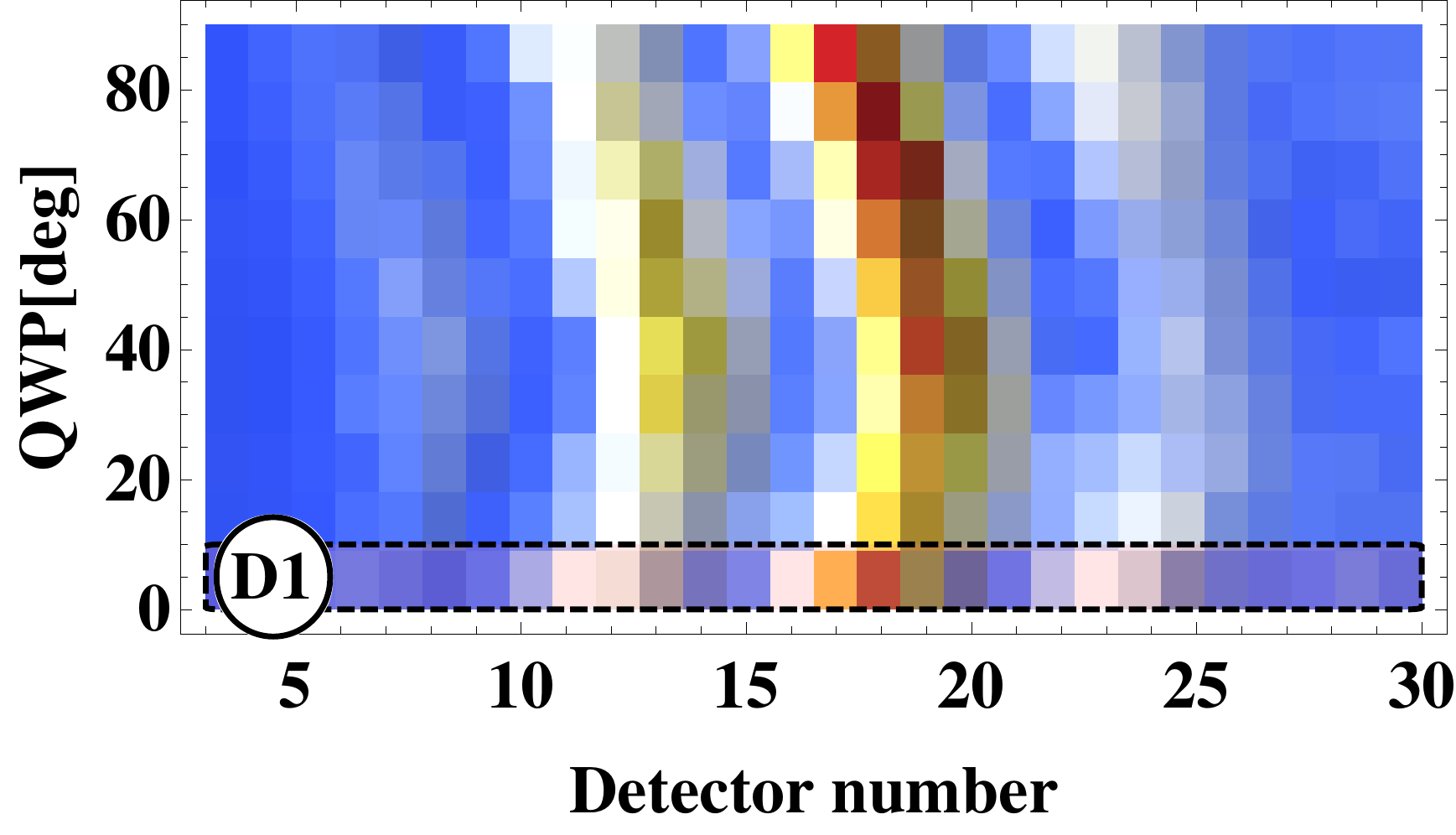}
%\put(15,35){\includegraphics[width=0.25\columnwidth]{bd-rot-bloch}}
%0.926199,0.909747,0.747788,0.839827,0.898374,0.95572,0.937743,0.972656,0.925,0.943463,0.929577
\end{overpic}}}
&
\multirow{2}{*}{\subfigure[]{\begin{overpic}[width=0.28\columnwidth]{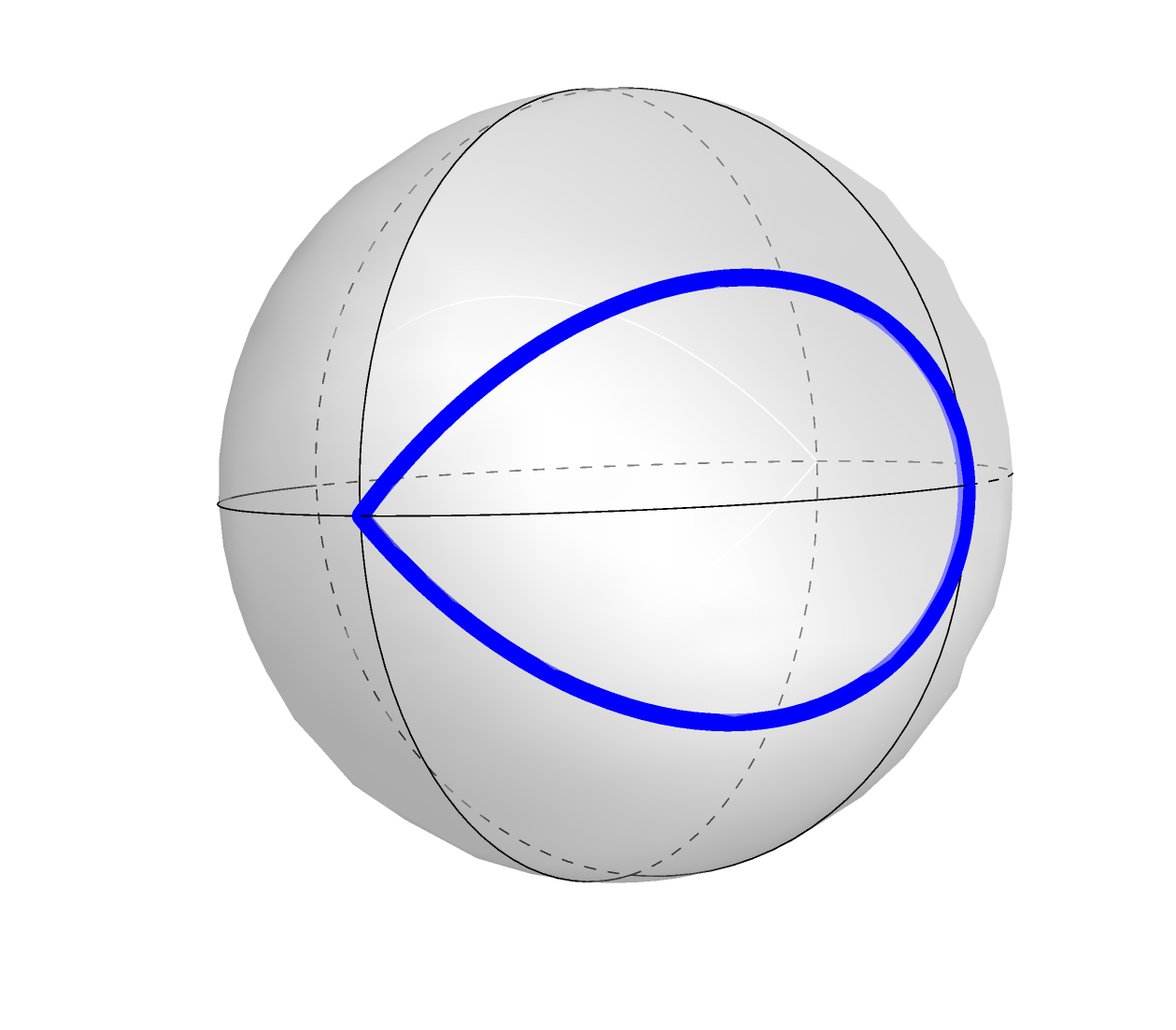}
\put(45,85){\tiny $\ket{H}$ }
\put(45,0){\tiny $\ket{V}$ }
\put(60,50){\tiny $\ket{\circlearrowleft}$ }
\put(24,35){\tiny $\ket{\circlearrowright}$ }
\put(89,40){\tiny $\ket{D}$ }
\put(5,47){\tiny $\ket{A}$ }
\end{overpic}}}
\\
&
\subfigure[\ heralded by D2]{\begin{overpic}[width=0.58\columnwidth]{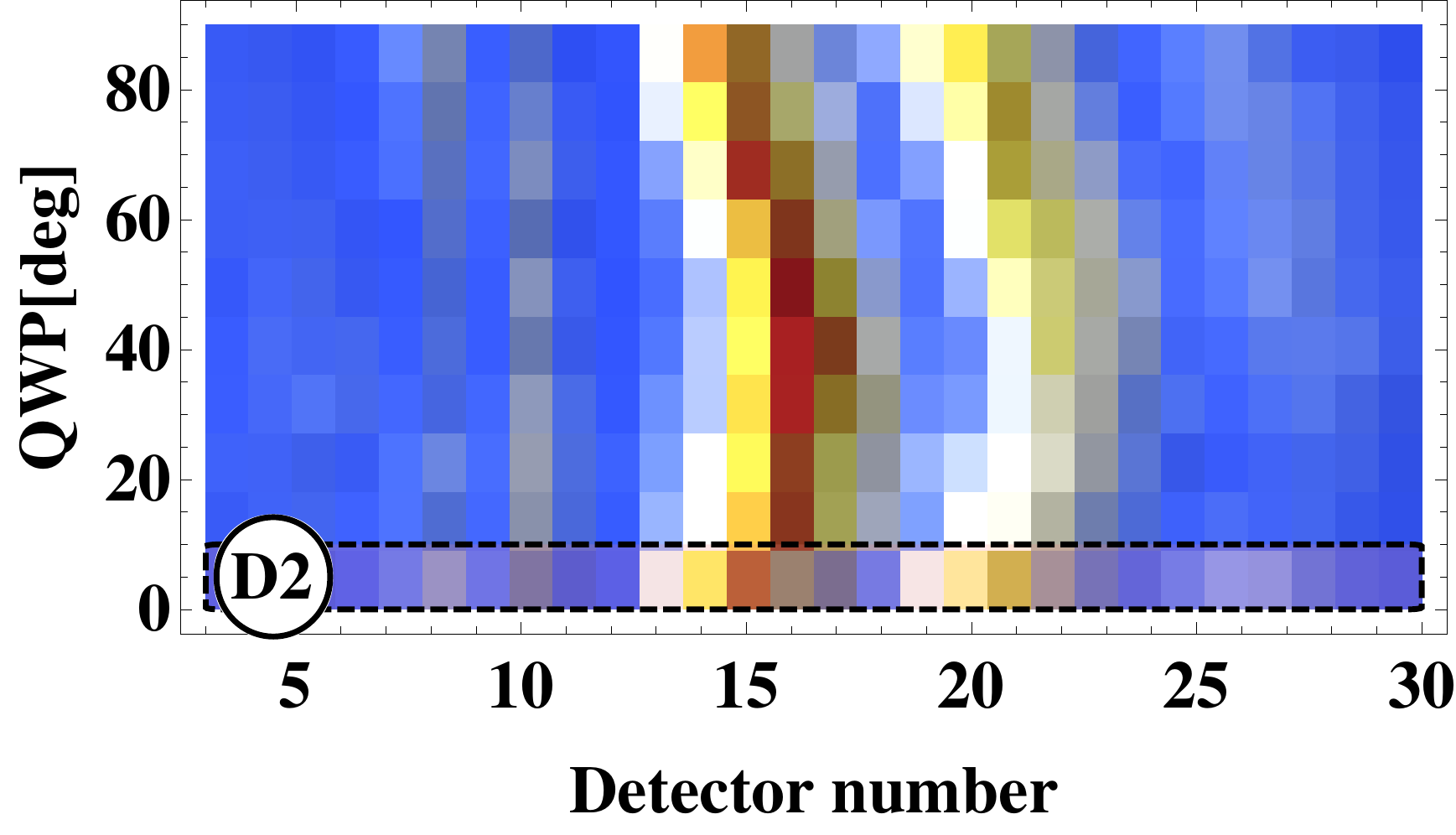}
%0.954704,0.939394,0.892256,0.853242,0.874194,0.92926,0.942492,0.922034,0.937705,0.974729,0.95572
\end{overpic}} 
&
\subfigure[]{\begin{overpic}[width=0.28\columnwidth]{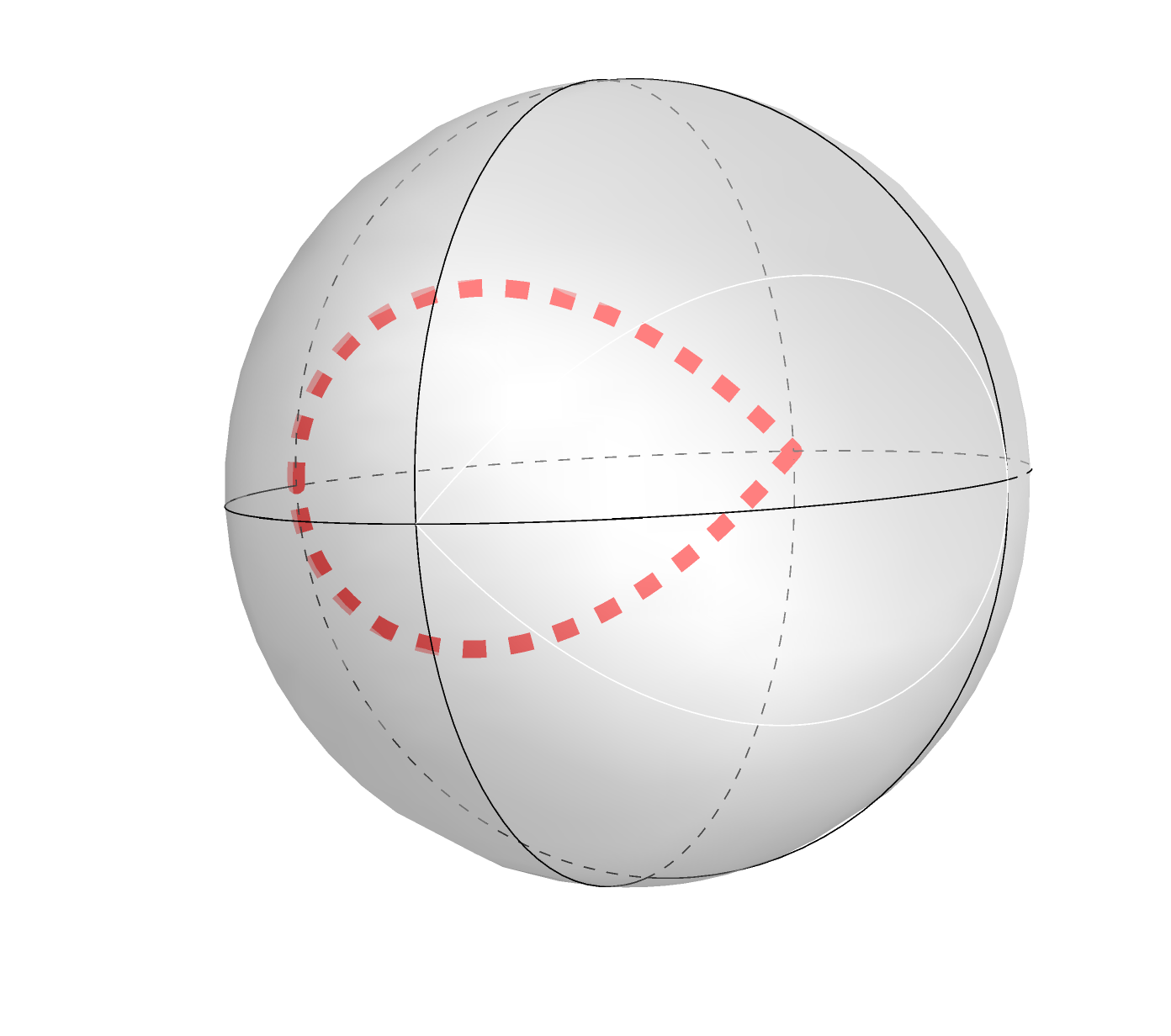}
\put(45,85){\tiny $\ket{H}$ }
\put(45,0){\tiny $\ket{V}$ }
\put(60,50){\tiny $\ket{\circlearrowleft}$ }
\put(24,35){\tiny $\ket{\circlearrowright}$ }
\put(89,40){\tiny $\ket{D}$ }
\put(5,47){\tiny $\ket{A}$ }
\end{overpic}}
\end{tabular}
\caption{  Interference. (A) The round (square) points show the interference pattern of the first $2000$ photons heralded by D2 (D1). The triangular points show the envelope that results from heralding by either polarization. Limitations of electronics resulted in fewer coincidences at detectors 7 and 10. (B,D) Interference pattern fringes move as the phase is changed remotely by the QWP.  The measurements are taken every $10$-degree rotation. See Supplementary Table I for the  visibilities of each set of measurements. (C,E) The trajectory of the Bloch vector related to the remotely prepared states heralded by (C) D1 and (E) D2. }
\label{fig:interf:rotations}
\end{figure*}

%likelihood ratio test
In order to compare the two methods, we perform a likelihood ratio test. This test tells us which model is better at reproducing the observed data (see discussion in Supplementary Information). First, we calculate the probability distribution of photon detections based on quantum mechanics. We then numerically simulate the corpuscular model $2.6 \times 10^6$ times using the best algorithm in Ref.~\cite{Jin2010a} to obtain its detection probability distribution, which is dependent on the number of detected photons. In contrast, the quantum mechanical distribution has no such dependence.  Next, we calculate how likely it is that our experimental data emerges from these probability distributions and compare them using the likelihood ratio, $\Lambda$. This test is independent of the numerical value of the ratio; as long as $\log \Lambda>0$, we can say that quantum mechanics is more accurate than the corpuscular model. Since $\log \Lambda \geq 0.83$ for all points in \figref{fig:cod:spdc}(B), we conclude that quantum mechanics is a better indicator of the behaviour seen in nature.

%comp fringes

In a second experiment, we use the same setup and generate polarization-entangled photons in the state $\ket{\psi} = \frac{1}{\sqrt{2}}(\ket{VH}_{s,i}+\ket{HV}_{s,i})$ with fidelity 0.94. Here $s,i$ represent the signal and idler photons. The orthogonal polarization states of the $842$ nm signal photon, $\ket{H}$ and $\ket{V}$, are transformed into the spatial states $\ket{\uparrow}$ and $\ket{\downarrow}$ by the calcite crystal. These refer to the two possible paths through the beam displacer. The resulting entangled state is $\ket{\psi} = \frac{1}{\sqrt{2}}(\ket{\downarrow H}_{s,i}+\ket{\uparrow V}_{s,i})$. The $776$ nm idler photon is sent to a polarization analyzer, which consists of waveplates, a polarizing beamsplitter and two detectors (see \figref{fig:setup:spdc}). The orientation of the HWP is set such that detection by D1 and D2 correspond to projection on $(\ket{H}+\ket{V})/\sqrt{2}$ and $(\ket{H}-\ket{V})/\sqrt{2}$, respectively. 

After taking data for 60s, we filter the detection events by choosing detections at either D1 or D2 as the trigger. If we choose D1 as the trigger, we herald the state $(\ket{\uparrow}+\ket{\downarrow})/\sqrt{2}$, which leads to the interference fringes shown in \figref{fig:interf:rotations}(A). Similarily, triggering by detection at D2 heralds $(\ket{\uparrow}-\ket{\downarrow})/\sqrt{2}$, resulting in a complementary interference pattern. The fringes are complementary because of the phase difference between the states heralded by D1 and D2. If we instead choose to herald using D1 \emph{or} D2 without distinguishing between the two, there is no interference pattern. This is because we effectively ignore the polarization state of the trigger photon, leaving the signal photon in a mixed state.
 
Because the photons are entangled, the phase of the interference pattern is correlated with the polarization state of the signal photon. To show that we indeed have entanglement between spatial and polarization degrees of freedom, we rotate the QWP in the polarization analyser. The resulting effect on the fringes are shown in \figref{fig:interf:rotations} (B,C). The phase of the pattern is clearly dependent on the polarization state of the trigger photon. In contrast, the polarization state of the trigger photon would have no effect on the phase of the interference pattern if these were non-entangled pairs. This heralding can also work in reverse. By post-selecting on a particular point in the interference pattern, it is possible to prepare the idler photon in a specific polarization state. Such a flexible remote state preparation could be very helpful in photonic quantum information processing.

% Conclusions
The double-slit experiment, which is at the ``heart of quantum mechanics'', has played a central role in our understanding and interpretation of quantum theory \cite{Feynman1965}. Now, over two hundred years after the first experiments by Thomas Young \cite{Young1802,Young1804}, our results provide the most complete picture of single-photon interference to date. Additionally, our time-resolved measurement techniques will dramatically decrease the difficulty of directly measuring the wave function of a system by performing weak measurements \cite{Kocsis2011,Lundeen2011}.  It will also allow us to herald a variety of polarization states in a multiplexed fashion, as well as facilitate the encoding and transfer of information using the hyper-entanglement of the spatial, temporal and polarization degrees of freedom \cite{Barreiro2005,Fickler2012,Kolenderski2011}.

\subsection{Acknowledgements}
The authors  acknowledge  funding from NSERC, Ontario Ministry of Research and Innovation (ERA program), CIFAR, Industry Canada and the CFI. The research leading to these results has received funding from the European Union Seventh Framework Programme (FP7/2007-2013) under grant agreement no.~257646. The authors thank Simone Bellisai and Franco Zappa for their contribution in developing the detectors array and Christopher Granade for insightful discussions. PK acknowledges support by Mobility Plus project financed by Polish Ministry of Science and Higher Education.

\subsection{Methods}
{\bf Experimental setup.} The details of the Sagnac-type source of photon pairs are described in Ref.~\cite{Hamel2010}, with a few modifications.  The pump is a 404 nm laser diode (Toptica Bluemode), and the down conversion crystal is a 30 mm PPKTP crystal phasematched to produce photons at $776$ and $842$ nm. The output of C1 has beam waist 1.3 mm. The calcite crystal is $41$ mm long, and the compensation crystal is $5$ mm long. Lens L1 is plano-convex (f=$150$ mm), lens L2 is aspherical (f=$11$ mm) and lens L3 is a plano-convex cylindrical  (f=25 mm). D1 and D2 are Perkin Elmer SPCM-AQ4C single photon detectors. The photon source produced around $2\times 10^6$ photon pairs/second which resulted in around $36\times 10^4$ fiber coupled pairs/second. Then the transmission of the calcite system decreased this number to approximately $72\times 10^3$, which results in around $2000$ detected coincidences/second. The SPAD array detector dark count rate gives rise to approximately $5$ accidental coincidences/second. All 32 channels of the SPAD array are recorded individually as time tags by two logic units (UQDevices).\
\\
{\bf SPAD array.} The SPAD array is a 32x1 array of single-photon avalanche diodes \cite{Guerrieri2010a,Tisa2007}, with pixel pitch of $100~\mu$m and photon detection efficiency $5$~\% in the range $770$ - $840$ nm. It has active area diameter of $50$ $\mu$m and a dark count rate of $100$ counts/s per pixel. For technical reasons, we use $28$ of the pixels.\\

%
%\bibliography{/media/Sklad/Piotr/publikacje/PKbase.bib}

\newpage
\pagebreak
\appendix
\section{Supplementary information}

Movie can be find here:\\ \mbox{\url{http://youtu.be/H11hJWIcUY0}}

\subsection{Additional measurements -- Coherent source}
\begin{figure}[h]
{\begin{overpic}[width=\columnwidth,trim = 0mm 28mm 5mm 10mm, clip]{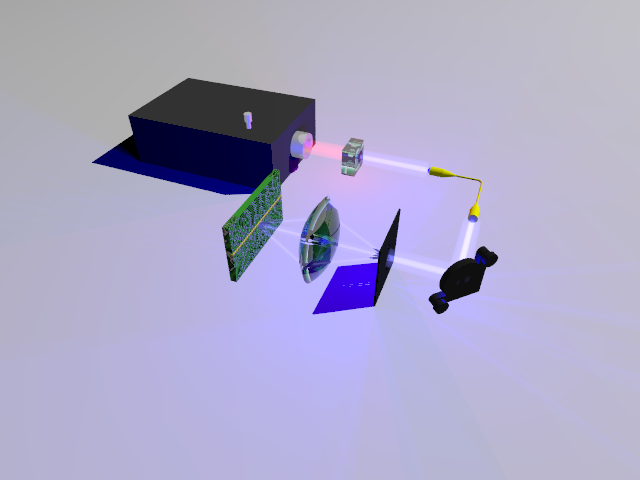}
\put(31,10){\scriptsize SPAD}
\put(31.03,7.25){\scriptsize Array}
\put(45,10.25){\scriptsize L3}
\put(55.5,7.5){\scriptsize Slits}
\put(74,8.75){\scriptsize M}
\put(77.05,23){\scriptsize L2}
\put(66,34.5){\scriptsize L1}
\put(52.25,38){\scriptsize BBO}
\put(26.5,48){\scriptsize Ti:Sapphire}
\end{overpic}}
\caption{Attenuated coherent states.  The Ti:Sapphire laser pumps the BBO. Photons are coupled into single-mode fiber using lens L1 and collimated by L2 before passing through the double slits. Lens L3 focuses the beam on the SPAD array.}
\label{fig:setup:coh}
\end{figure}
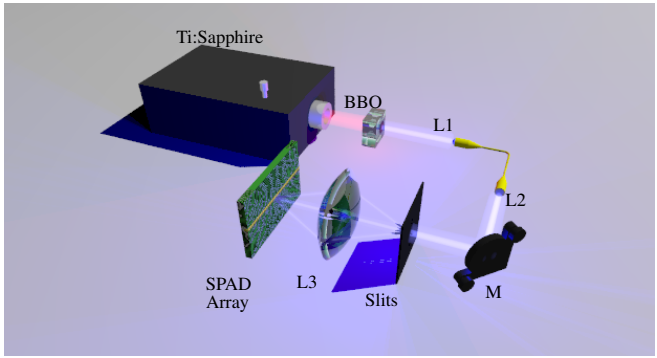

Another experimental setup depicted in \figref{fig:setup:coh} has three components: an attenuated laser as a single photon source, a slit system \cite{Kolenderski2011} and a SPAD array detector   \cite{Guerrieri2010a,Tisa2007}.

The attenuated coherent states experiment is based on fiber-coupled frequency-doubled 1 ps pulses at $396$ nm attenuated such that the SPAD array detects approximately $200$ photons/s. This results in $0.06$ average photons per pulse which, assuming Poissonian statistics, makes the probability of more than one photon arriving at the slit at the same time negligible.

The Ti:Sapphire laser outputs $792$ nm pulses, and is used to pump a $2$ mm BBO crystal. Single mode fiber and collimating lens L2 output a gaussian beam with a radius of $0.85$ mm (FWHM). Lens L3 has a focal length of $10$ cm. The slits are $500 \mu$m high, $30 \mu$m wide and separated by $100$ $ \mu$m. 

Lens L2 is chosen such that the impinging photons' spatial mode size is much larger than the slits' characteristic size. This results in a uniform illumination and transmission of $4 \%$.

After passing through the slits, the photon propagates through lens L3 and is detected by a SPAD array in the focal plane, where the interference pattern is formed. The photons' spatial and temporal modes, in addition to the optics, are chosen such that the interference pattern's characteristic size is comparable to the dimensions of half of the array. This, in conjunction with the high quantum efficiency of the SPAD array at this wavelength, which is $40\%$, allows an optimal signal-to-noise ratio at the single-photon level. It also allows the expected interference pattern minima (maxima) to coincide with odd (even) numbered SPAD array pixels.

The timing information of a detected photon is recorded only if it is coincidental with a reference pulse from the Ti:Sapphire pulsed laser, thus reducing background noise from dark counts and stray light. The timing resolution of the electronics and SPAD array result in a detection window of $312$ ps and an effective $6$ total dark counts/s.

\begin{figure}
\begin{tabular}{c c}
\multirow{3}{*}{\subfigure[$N=200$, $t=1$ s]{\includegraphics[width=0.45\columnwidth]{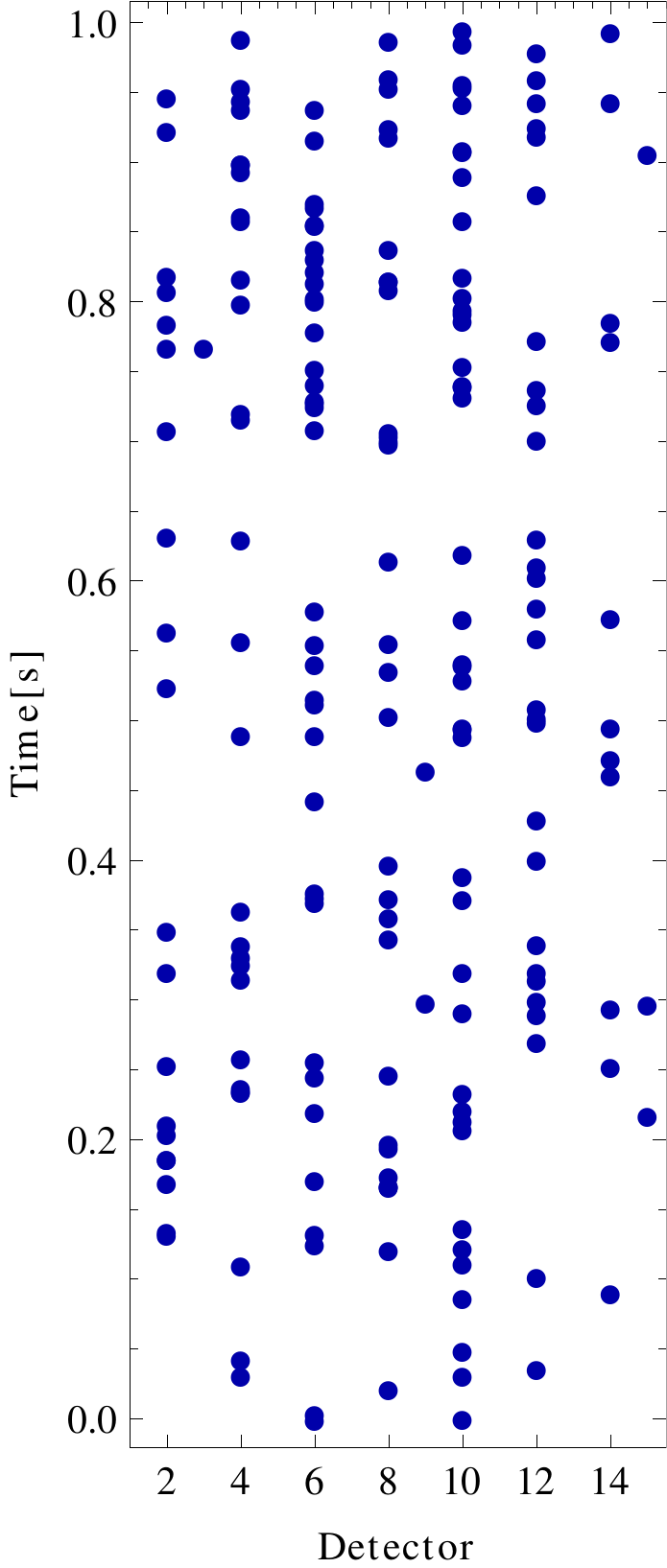}}} &  \vspace{2.8cm}
\multirow{2}{*}{\subfigure[$N=2000$, $t=9.95$ s]{\includegraphics[width=0.43\columnwidth]{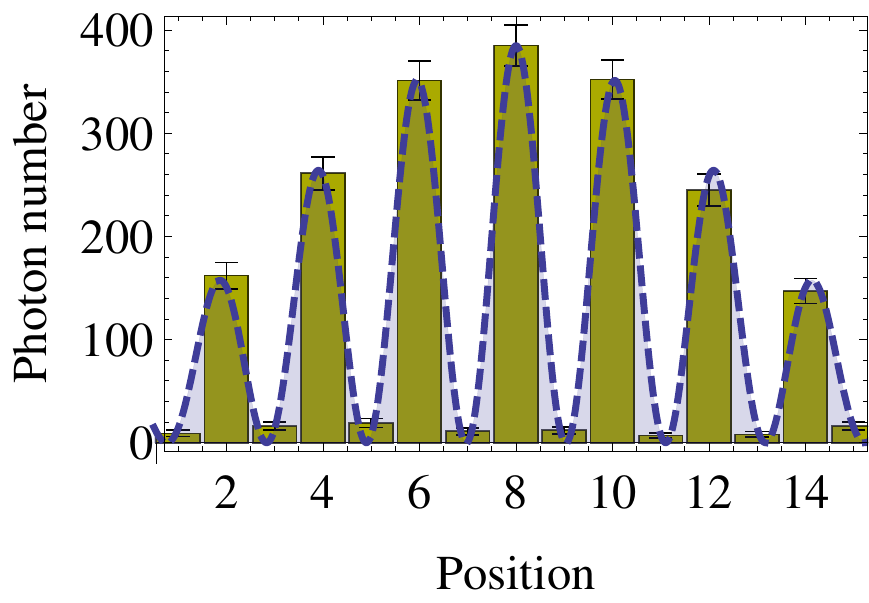}}}  \\
&\subfigure[$N=200$, $t=1$ s]{\includegraphics[width=0.43\columnwidth]{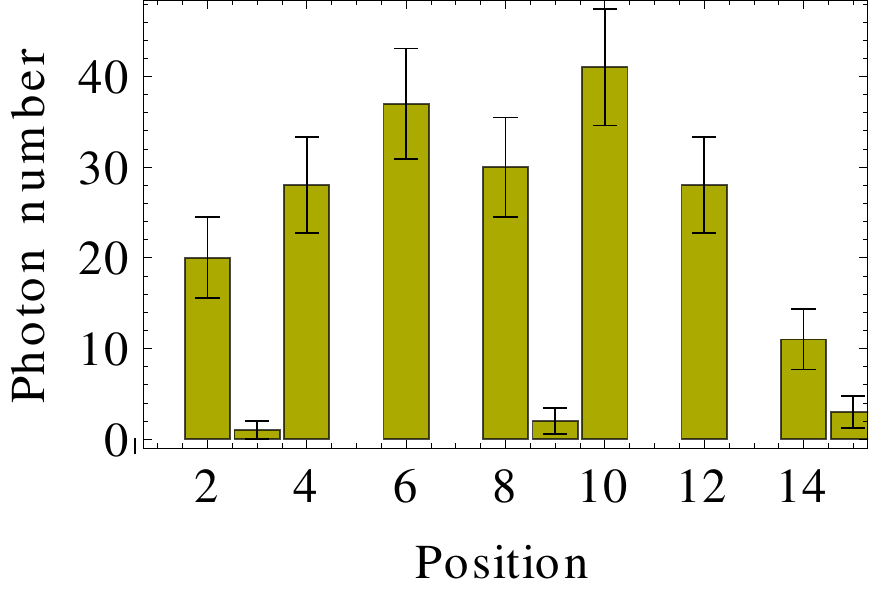}}  \\
&\subfigure[$N=20$, $t=0.14$ s]{\includegraphics[width=0.43\columnwidth]{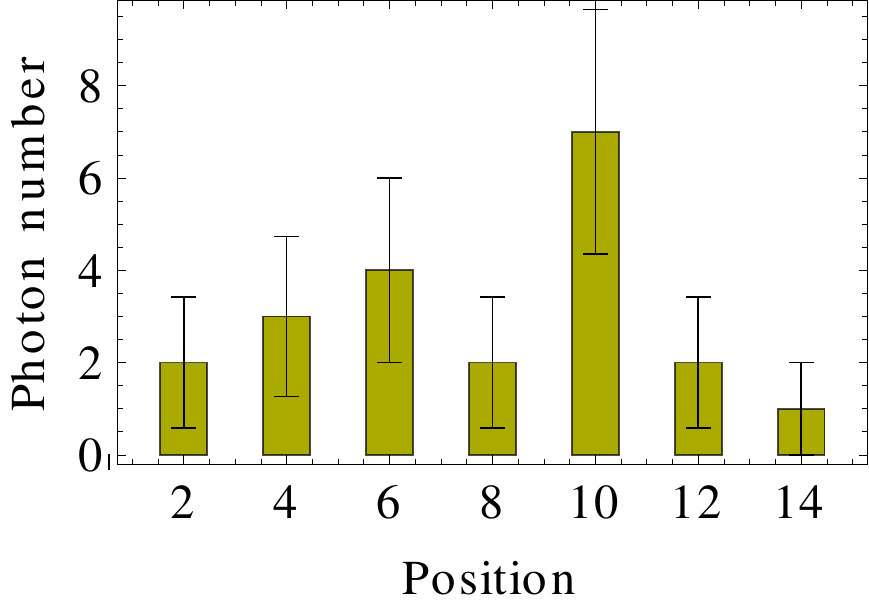}}  
\end{tabular}
\caption{Interference pattern buildup for 2 slits. (a) First $200$ detections in time and (b-d) statistics of first $20$, $200$ and $2000$ detections are presented.}
\label{fig:reel:coh}
\end{figure}

We observe the buildup of the interference pattern in time, as seen in \figref{fig:reel:coh}. Note that odd-numbered SPAD pixels detect significantly fewer photons than even-numbered pixels, which correspond to the minima and maxima of the interference pattern. The contrast between the neighbouring pixels is clearly visible after only $20$ photon detections, indicating the existence of some sort of interference pattern from the beginning of the measurement.

%The coefficient of determination, $R^2$, as a function of the number of detected photons is depicted in \figref{fig:cod:coh}. 

%\begin{figure}[h]
%{\includegraphics[width=0.8\columnwidth]{bd-interf-rotA-ch15-R2-belt}}
%%\includegraphics[width=0.95\columnwidth]{bd-interf-R2-dlm}
%\caption{ Coefficient of determination, $R^2$, as a function of detected photons number for attenuated coherent states. Dots represent the fit for experimental data. For a given photon number, the statistics of $R^2$ was generated after $10^5$ numerical event-by-event and quantum mechanical (QM) model simulations. The belts  show $50$\% of the most frequent values of $R^2$.}
%\label{fig:cod:coh}
%\end{figure}

%\subsection{Triple slit experiment}

\begin{figure}
\centering
\begin{tabular}{c c}
%s111reel
\multirow{3}{*}{\subfigure[$N=200$, $t=894 \mu$s]{\includegraphics[width=0.45\columnwidth]{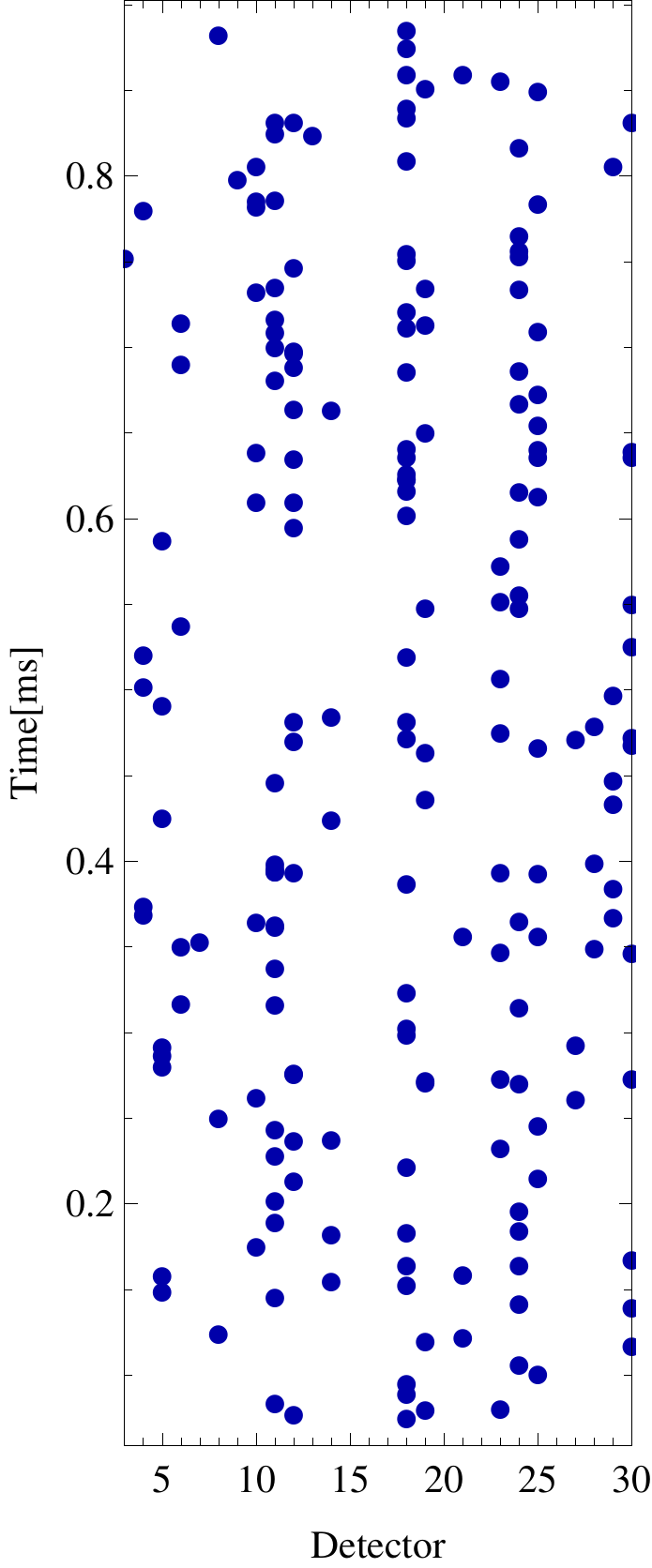}}} & \vspace{2.8cm} \multirow{2}{*}{\subfigure[$N=2000$, $t=9.88$ ms]{\includegraphics[width=0.43\columnwidth]{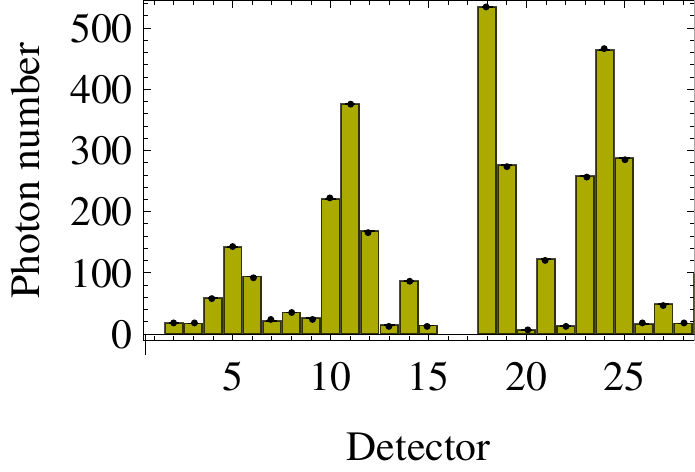}}} \\
&\subfigure[$N=200$, $t=894 \mu$s]{\includegraphics[width=0.43\columnwidth]{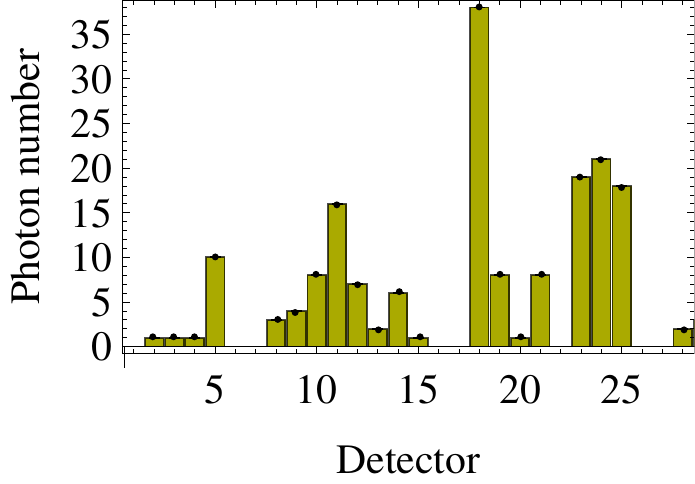}} \\
&\subfigure[$N=20$, $t=74 \mu$s]{\includegraphics[width=0.43\columnwidth]{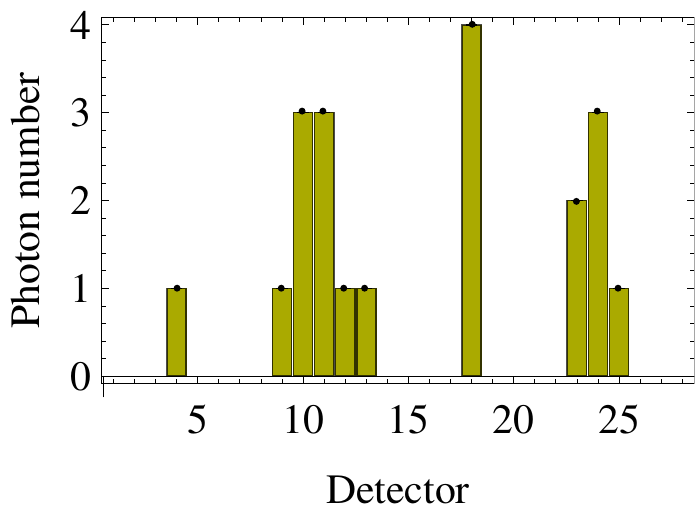}} \\
\end{tabular}
\caption{Interference pattern buildup for 3 slits. (a) First $200$ detections in time and (b-d) statistics of first $20$, $200$ and $2000$ detections are presented. Note that due to technical problems, pixels 16 and 17 were disconnected.}
\label{fig:hist:111}
\end{figure}

Additional measurements are done for three slits using attenuated coherent states.  The imaging optics are adjusted to fit the SPAD array and interference pattern characteristic dimensions. The consecutive detection events as a function of detector number and time, as well as the histograms of the recorded time tags, are depicted in \figref{fig:hist:111}.

\subsection{Interference pattern analysis}

After passing through the system of a birefringent crystal and polarizer, a wave is in superposition of two spatial gaussian modes displaced by the distance $d$. Assuming a characteristic radius of $w$, the corresponding mode functions are:
\begin{equation}
u_\pm(x,z=0) \propto \exp\left(\frac{-(x\pm d/2)^2}{w^2}\right).
\label{eq:basis}
\end{equation}
Note that these modes are nearly orthogonal if the displacement $d$ is sufficiently large. In our case, $w=1.4$ mm and $d=3.68$ mm and the overlap is ${ \sqrt{\frac{2}{\pi w^2}}e^{-\frac{d^2}{2 w^2}}}=0.0096$. 

Next, the standard Fresnel propagation allows to compute the filed in the focal plane of the lens:
\begin{eqnarray}
 		u_\pm(x,z=f) \propto e^{-\frac{\pi  x \left(\pi  w^2 x\pm 2 i d f \lambda \right)}{f^2 \lambda ^2}}.
\end{eqnarray}
By modification of the relative amplitudes and the phases of the two modes, one can prepare any superposition of the following form:
\begin{equation}
\alpha_+ u_+(x,z) + \alpha_- u_-(x,z).
\label{eq:qubit}
\end{equation}
Equations (2) and (3) allow us to predict the interference patten shape and its dependence on the input state. 

\section{The corpuscular model}

The corpuscular model of the double-slit experiment \cite{Jin2010a} gives an alternative description for the buildup of the interference pattern. The detectors are based on deterministic learning machines, whose internal states are updated with each ''messenger'' (photon) detection. The messengers propagate for a specific time after passing through the slits, acquiring a phase $\phi$, which then updates the detectors' states according to the equations \cite{Jin2010a}:
\begin{eqnarray}
\mu_{k-1}&=&\gamma(1-w_{k-1}),\\
p_k&=&\mu_{k-1}p_{k-1}+(1-\mu_{k-1})e_k,\\
w_k&=&\kappa w_{k-1}+(1-\kappa)\frac{||p_k-p_{k-1}||}{2},
\end{eqnarray}
where $\kappa$ and $\gamma$ are constants associated with the detectors, $p_k$ is a parameter that is updated with each detection, and $w_k$ is the internal state of the detector.

In addition to the likelihood ration test presented in \figref{fig:cod:spdc}, we comment on few observations on the corpuscular model based on the numerical simulations and our measurements. 

We compare the this model to the other aspects of our experiment, including complementary fringes and the shifting interference patterns. It is very clearly shown in \figref{fig:interf:rotations}(a) that the two sets of fringes are extracted from the same measurement data. This situation would confuse the detectors' learning process, thus telling us that entanglement resides outside of the scope of the corpuscular model.

\section{Interference pattern visibilities}
\begin{table}[h]
\begin{tabular}{c  c  c }
QWP [deg] & Vis., [$\%$], D1& Vis.,[ $\%$], D2\\
\hline \hline
0 & 93$\pm2$ & 96$\pm2$\\ 
10 & 94$\pm2$ & 98$\pm2$  \\ 
20 & 93$\pm2$ & 94$\pm2$ \\ 
30 & 97$\pm2$ & 92$\pm2$ \\ 
40 & 94$\pm2$ &94$\pm2$ \\ 
50 & 96$\pm2$& 93$\pm2$ \\ 
60 & 90$\pm2$ &87$\pm2$ \\ 
70 & 84$\pm2$  &85$\pm2$ \\ 
80 & 75$\pm2$  &89$\pm2$ \\ 
90 & 91$\pm2$  &94$\pm2$\\ 
100 & 93$\pm2$ &95$\pm2$ \\ 
\end{tabular}
\caption{Visibilities for patterns obtained by QWP rotations, see \figref{fig:interf:rotations}.}
\label{TableVis}
\end{table}

\section{Likelihood ratio test}

The likelihood ratio test \cite{Casella2001} allows to qualitatively evaluate which of the two models is a better predictor of the measured data. This can be done by looking at the ratio, $\Lambda$ of the probability, $P(D|\text{M1})$, to get a certain set of data, D, under one model, M1, to the probability, $P(D|\text{M2})$ to get the same data, D, under a second model, M2.  It is convenient to take the logarithm of this ratio:
\begin{eqnarray}
\log \Lambda=\log P(D|\text{M1})-\log P(D|\text{M2}),
\end{eqnarray}
where $D=\{k_1, k_2, ..., k_{28}\}$ in our case represents the photon counts distribution measured by the SPAD array ($k_i$, stands for counts of $i$th detector) and $\text{M}=\{p_1, p_2,...,p_{28}\}$ is the probability distribution given by a model ( $p_i$ is the probability of detecting a photon by $i$th detector). If $\log \Lambda>0$, we can say that M1 describes experimental data better than M2 does.

The probability to get a distribution, D, assuming probability distribution, M, is calculated using a multinomial expansion:
\begin{eqnarray}
P(D|\text{M})=\sum_{k_1+k_2+...+k_{28}} \frac{n!}{{k_1}!{k_2}!...{k_{28}}!} \prod_{1\leq n \leq28}^{} p_n^{k_n}.
\end{eqnarray}

The probability distribution for the quantum mechanics, M1, is computed in the following way. We fit the photon statistics acquired after detection of $98000$ photons to the quantum mechanical model. The fitted parameters are intensity, transverse shift and magnification. The remaining setup parameters were fixed to the measured values. The coefficient $R^2$ is  $0.99$. Note that within this model the probability distribution does not depend on detected photon number.

This is the feature that differs quantum mechanics and the corpuscular model. To calculate the photon detection probability distribution, M2, for the corpuscular theory, $2.6\times 10^6$ simulations were made for photon number in the range $1\dots 2000$. In contrast to quantum mechanics, the probability distribution for this model depends on the number of photons that have been detected. 

It is important to note that it is meaningless to compare the values of the likelihood ratio for different sets of data, e.g. different numbers of photons.

\end{document}